\documentclass[11pt]{article}
\usepackage{hyperref}
\pdfoutput=1

\begin{document}
\title{Droplet coalescence and breakup: Numerical simulations using moving mesh interface tracking with dynamic mesh adaptation}
\author{Shaoping Quan and Jing Lou\\
        \vspace{6pt}
        Institute of High Performance Computing, Singapore $138632$}

\maketitle

\begin{abstract}
Droplet(s) breakup and coalescence have been simulated by the finite volume/moving mesh interface tracking method (MMIT) with adaptive mesh refining and coarsening. In this method, the interface is zero-thickness and moves in a Lagrangian fashion. The breakup and the coalescence of the interfaces are simulated by the mesh separation and mesh combination schemes. Three cases are displayed in this video, which include the relaxation and breakup of an initially extended droplet, two identical droplets off-center collision, and a droplet pair in a shear flow. The mesh adaptation is capable of capturing large deformations and thin regions, and the smallest length resolved is three orders of magnitude smaller than the droplet radius. The fluid dynamics videos of the simulations are to be presented in the Gallery of Fluid Motion, $2010$.
\end{abstract}

\section{Introduction}
The two videos for numerical simulations of droplet(s) breakup and coalescence are
\href{http://ecommons.library.cornell.edu/bitstream/1813/8237/2/LIFTED_H2_EMST_FUEL.mpg}{Video1}
and
\href{http://ecommons.library.cornell.edu/bitstream/1813/8237/4/LIFTED_H2_IEM_FUEL.mpg}{Video2}.

In the link videos, the simulations are performed using a finite volume staggered mesh method coupled with moving mesh interface tracking (MMIT) scheme to locate the interface \cite{Quan_Schmidt_2007}. The advantages of this method is: 1) The interface is tracked by the triangular surfaces of the interior tetrahedral mesh elements, and thus zero-thickness; 2) The jumps of the fluids' properties and stresses across the interface are implemented directly without smoothing; 3) The mass of each phase is naturally conserved due to the lagrangian motion of the interface. To capture the large deformations and wide range of length scales, a number of mesh adaptation schemes are implemented, and these mesh adaptation schemes distribute the mesh in an optimized way to achieve computing efficiency. The topological changes in multiphase flows are computed by the mesh separation and mesh combination schemes \cite{Quan_Lou_2009}. 

Three cases are shown in the videos. The first case is the simulation of the relaxation and breakup of an initially extended droplet. The initial drop has a shape with a cylinder in the center and two bulbous ends. The surface forces are not balanced and thus the droplet experience a contraction and for this case the droplet eventually breaks up to three satellited drops. The density and viscosity ratios are 10, and 0.1, respectively, and the Ohnesorge number is 0.037. The result was reported previously \cite{Quan_Schmidt_2009}, however, for the current simulation, the smallest neck radius is 10 times smaller than the previous simulation. The second case is an off-center droplet pair collision, and due to the large initial momentum, the two droplets coalesce and finally break up into five satellite drops. This case has a Weber number of 32, a Reynolds number of 80, and both the density and viscosity ratios of 10. The detail of the simulation can be found in \cite{Quan_Lou_2009}. The third case is a droplet pair in a shear flow with matched densities of the two fluids, and a viscosity ratio of 1.4, a Reynolds number of 0.1, and a Capillary number of 0.14. Initially the droplets are 5$R_0$ ($R_0$ is the radius of the spherical drop) away from each other streamwisely, and have a separation of 0.5$R_0$ in the direction perpendicular to the free stream direction. The capability of our schemes in dealing with topological changes and resolving multiple length scales has been demonstrated.


\begin{thebibliography}{3}

\bibitem{Quan_Lou_2009}
S.~P. Quan, J.~Lou, and D.~P. Schmidt.
\newblock Modeling merging and breakup in the moving mesh interface tracking
  method for multiphase flow simulations.
\newblock {\em J. Comput. Phys.}, 228:2660--2675, 2009.

\bibitem{Quan_Schmidt_2007}
S.~P. Quan and D.~P. Schmidt.
\newblock A moving mesh interface tracking method for 3d incompressible
  two-phase flows.
\newblock {\em J. Comput. Phys.}, 221:761--780, 2007.

\bibitem{Quan_Schmidt_2009}
S.~P. Quan, D.~P. Schmidt, J.~S. Hua, and J.~Lou.
\newblock A numerical study of the relaxation and breakup of an elongated drop
  in a viscous liquid.
\newblock {\em J. Fluid Mech.}, 640:235--264, 2009.

\end{thebibliography}

\end{document}